\documentclass[aps,prb,showpacs,twocolumn,superscriptaddress]{revtex4-1}

\usepackage{graphicx}
\usepackage{amsmath,amssymb}
\usepackage{hyperref}
\usepackage{natbib}
\usepackage{bm}

\begin{document}

%
\title{Topological invariants in interacting Quantum Spin Hall: a Cluster Perturbation Theory  approach.}

\date{\today}
\author{F.~Grandi}
\affiliation{Dipartimento di Scienze Fisiche, Informatiche e Matematiche, Universit\`a di Modena e Reggio Emilia, Via Campi 213/A, I-41125 Modena, Italy }
\author{F.~Manghi}
\affiliation{Dipartimento di Scienze Fisiche, Informatiche e Matematiche, Universit\`a di Modena e Reggio Emilia, Via Campi 213/A, I-41125 Modena, Italy }
\affiliation{CNR - Institute of NanoSciences - S3}

\author{O.~Corradini}
\affiliation{Facultad de Ciencias en F\'isica y Matem\'aticas, Universidad Aut\'onoma de Chiapas, Ciudad Universitaria, Tuxtla Guti\'errez 29050, M\'exico}
\affiliation{Dipartimento di Scienze Fisiche, Informatiche e Matematiche, Universit\`a di Modena e Reggio Emilia, Via Campi 213/A, I-41125 Modena, Italy }

\author{C.M.~Bertoni}
\affiliation{Dipartimento di Scienze Fisiche, Informatiche e Matematiche, Universit\`a di Modena e Reggio Emilia, Via Campi 213/A, I-41125 Modena, Italy }
\affiliation{CNR - Institute of NanoSciences - S3}
\author{A.~Bonini}
\affiliation{Dipartimento di Fisica e Astronomia, Universit\`a di Bologna, Via Irnerio 46, I-40126 Bologna, Italy }

\begin{abstract}

Using Cluster Perturbation Theory  we calculate Green's functions, quasi-particle energies and  topological invariants for interacting electrons on a 2-D honeycomb lattice, with intrinsic spin-orbit coupling and on-site e-e interaction. This allows to define the parameter range (Hubbard U vs spin-orbit coupling) where the 2D system behaves as a  trivial insulator or   Quantum Spin Hall insulator. This behavior is confirmed by the existence of gapless quasi-particle states in honeycomb ribbons. 
We have discussed the importance of the cluster symmetry and the effects of   the lack of full translation symmetry typical of CPT and of most  Quantum Cluster approaches. Comments on the limits of applicability of the method are also provided.
\end{abstract}
\maketitle


Topological invariants are by now widely recognized as a powerful tool to characterize different phases of matter; in particular they turn out to be  useful in the classification of topological insulators.~\cite{RevModPhys.82.3045,Ando} In the topological insulator phase, solids are characterized by gapped bulk bands but present gapless edge states that allow charge or spin conductivity on the boundaries. The presence of such gapless edge states is linked to the emergence of non-vanishing topological invariants via a bulk-boundary correspondence.~\cite{PhysRevB.74.045125} This topological feature ensures the robustness of the edge states against disorder.~\cite{PhysRevB.73.045322,PhysRevLett.96.106401}

A two-dimensional honeycomb lattice with  spin-orbit coupling has been identified as a remarkable and paradigmatic example of topological insulator. This system is the prototype of the so-called Quantum Spin Hall (QSH) system presenting a quantized spin-Hall conductance at the boundaries. The topological nature of QSH insulators  is identified by  a time-reversal (${\cal T}$) - topological invariant $\mathbb{Z}_2$~\cite{PhysRevLett.95.146802,PhysRevB.74.195312,PhysRevB.76.045302}. In the same way as the Thouless-Kohmoto-Nightingale-den Nijs~\cite{Thouless:1982zz} (TKNN) topological invariant was defined for the integer quantum Hall effect, the above $\mathbb{Z}_2$ invariant was defined for the topological insulator  in terms of band eigenvectors and, as such, only applies to noninteracting systems.  On the other hand, in the presence of  electron-electron interaction, there are generalizations of the TKNN invariant based on twisted boundary conditions~\cite{PhysRevB.31.3372} and on many-body Green's functions,~\cite{PhysRevB.85.165126,PhysRevX.2.031008,PhysRevB.83.085426} and unlike the former, the latter construction can be straightforwardly extended to the $\mathbb{Z}_2$ invariant to classify interacting QSH systems.

The field of interacting topological insulators is attracting growing interest (see Refs.~\onlinecite{Assaadrev,doi:10.1142/S0217984914300014} for  recent reviews on two-dimensional systems) and the definition of theoretical and computational tools to evaluate topological invariants in the presence of e-e interaction is extremely timely.   The approach that seems most promising is the one developed in Ref.~\onlinecite{PhysRevX.2.031008} where it has been demonstrated that topological invariants are determined by the behaviour of the one-particle propagator at zero frequency only; more precisely it has been shown that the  eigenvectors of the single particle Hamiltonian that yields topological invariants for non-interacting systems, should be replaced in the interacting case by the  eigenvectors of the operator $G^{-1}(k,\omega)$ at $\omega=0$ and  momentum $k$. The extension of topological invariants to interacting systems is in this sense  straightforward, the only demanding task remaining the determination of the dressed Green's function. This concept has been recently applied to identify the topological character of heavy fermion mixed valence compounds~\cite{PhysRevB.88.035113,PhysRevLett.111.176404,PhysRevLett.110.096401,PhysRevB.87.085134} and of the half-filled honeycomb lattice with an additional  bond dimerization~\cite{PhysRevB.87.205101}.

In recent years a new class of many-body approaches  has been developed to calculate the one-particle Green's function of extended systems solving  the many body problem in a subsystem of finite size and  embedding it within an infinite medium. These methods gather under the name of Quantum Cluster theories~\cite{RevModPhysQC} and  include Cluster Perturbation Theory~\cite{Senechal,Senechal3} (CPT), Dynamical Cluster Approach~\cite{DCA} (DCA), Variational Cluser Approximation~\cite{VCA} (VCA), Cellular Dynamical Mean Field Theory~\cite{Senechal_CDMFT} (CDMFT). They  have found an unified language within the variational scheme  based on the the Self Energy Functional approach~\cite{SFA}. These methods, with different degrees of accuracy, give access to non trivial many body  effects and have been applied both to model systems  and to realistic solids~\cite{Eder,ManghiMnO}.

In this paper we consider the Kane-Mele-Hubbard model~\cite{PhysRevB.82.075106,PhysRevB.84.205121,PhysRevLett.106.100403,PhysRevB.85.115132} describing  a 2-dimensional honeycomb lattice with both  local e-e interaction and spin-orbit coupling and we adopt an approach based on CPT to determine the one-particle propagator, the \emph{topological hamiltonian} $G^{-1}(k,\omega=0)$ and its eigenvectors. This allows us to identify a general procedure that can be extended to any Quantum Cluster approach and to investigate how Green's function- based topological invariants can be effectively calculated.

The paper is organized as follows: in section~\ref{sec1} we recall  how topological invariants  can be obtained in terms of $G^{-1}(k,\omega=0)$. Section~\ref{sec2} describes  how topological invariants are obtained by CPT  and  section~\ref{sec3} reports the results in terms of topological invariants and spectral functions for the Kane-Mele-Hubbard model of   2D and 1D honeycomb lattices.

\section{Topological hamiltonian and topological invariants}
\label{sec1}
In the search for an extension of topological invariants from the non-interacting to the interacting case, the Green's function has proved to be the fundamental tool~\cite{PhysRevB.85.165126,PhysRevX.2.031008,PhysRevLett.105.256803,PhysRevB.83.085426}. As shown in Refs.~\onlinecite{PhysRevB.85.165126,PhysRevX.2.031008} the dressed one-particle Green's function at zero frequency contains all the topological information that is required to calculate topological invariants:  the inverse of the Green's function at zero frequency defines a fictitious  noninteracting topological hamiltonian \cite{0953-8984-25-15-155601}
\begin{align} h_{topo}(k)\equiv -G^{-1}(k,0)
\label{htopo}
\end{align}
and its eigenvectors
\begin{align}
\label{eigen}
h_{topo}(k)|k,n,s\rangle = \epsilon_{n s}(k)|k,n,s\rangle
\end{align}
are the quantities to be used to compute the topological invariants for the interacting system. Here $n$, $s$ are band and spin indices respectively ($s=\uparrow\downarrow$).  The latter is a good quantum number if --as in the model we study below-- the spin orbit interaction only involves the $z$  component of the spin.

Hence, we can take the time-reversal operator to be
\begin{align*}
\Theta = \mathbb{I} \otimes i\sigma_y\, K
\end{align*}
where $\sigma_y$ acts on the spin indices, $K$ denotes complex conjugation and $\mathbb{I}$ is the identity for the sublattice indices. The matrix
\begin{align}
\label{timerev}
w_{n s, n' s'}(k) \equiv \langle -k,n,s|\Theta|k,n',s'\rangle
\end{align}
is thus a block-diagonal matrix, and is antisymmetric at time-reversal invariant momenta (TRIM) $\Gamma_i$ defined by the condition that $-\Gamma_i= \Gamma_i +\mathcal{G}$ with $\mathcal{G}$ a reciprocal lattice vector. The generalized $\mathbb{Z}_2$ topological invariant can thus be defined \cite{PhysRevB.74.195312,PhysRevX.2.031008} as the exponent $\Delta$ in the expression

\begin{align}
(-1)^{\Delta} \equiv \prod_{TRIM}\frac{\sqrt{\det[w(\Gamma_i)]}}{{\rm Pf} [w(\Gamma_i)]}
\end{align}
and used to classify trivial insulators ($\Delta =0$, mod 2) from topological QSH insulators ($\Delta =1$, mod 2).
In the presence of inversion symmetry this definition is even simpler, involving just the parity eigenvalues $\eta_{n}(\Gamma_i)= \pm 1 $ of the occupied bands at $\Gamma_i$  for any of the two spin sectors
\begin{align}
\label{z2}
(-1)^{\Delta} =\prod_{TRIM} \prod_{n=1}^{N} \eta_{n}(\Gamma_i)~.
\end{align}

The  definition  of  $\mathbb{Z}_2$ for an  interacting   system is thus formally identical to the non-interacting case, involving in both cases the eigenstates of a single particle hamiltonian; in the presence of e-e interaction the difficult task remains the calculation of  the  topological hamiltonian in terms of the interacting Green's function.  In the next section we will describe how this can be done within the CPT paradigm.

\section{Kane-Mele-Hubbard model  and CPT}
\label{sec2}

We are interested in the Kane-Mele-Hubbard model for a 2D honeycomb lattice
\begin{align}
\label{kmh}
\hat{H} = \sum_{i l, i' l' s} t_{il,i'l'}(s)\hat{c}_{i l s}^{\dag} \hat{c}_{i' l'  s} +
U \sum_{i l} \hat{c}_{i l \uparrow}^{\dag}\hat{c}_{i l \uparrow}\hat{c}_{i l \downarrow}^{ \dag}\hat{c}_{i l  \downarrow}~.
\end{align}
The hopping term  $t_{il,i'l'}(s)$  includes both the  first-neighbor spin-independent hopping  and the Haldane-Kane-Mele
second-neighbor spin-orbit coupling~\cite{PhysRevLett.61.2015,PhysRevLett.95.146802} given by
$
\imath t_{KM} s_{z} ({d_{1}} \times {d_{2}})_{z}
$,
 where $d_{1}$ and $d_2$ are  unit vectors  along the two bonds that connect site  $i l$ with site $i' l'$. Here $i,i'$ run over the $ M$ atomic positions within the unit cell (cluster) and $l,l'$ refer to lattice vectors identifying  the unit cells of  the lattice.  The on-site e-e repulsion is described by the $U$-Hubbard term.

In order to solve the eigenvalue problem~\eqref{eigen}, in strict analogy with what is done in any standard Tight-Binding scheme for non-interacting hamiltonians, a Bloch basis  expression of the topological hamiltonian, namely of the dressed Green's function and of its inverse, is required
\begin{align}
\label{gij}
 G_{ij}(k, \omega)=\langle \Psi_0|\hat{c}_{k  i }^{\dag}\hat{G} \hat{c}_{k j}|\Psi_0 \rangle
 + \langle \Psi_0|\hat{c}_{k  i }\hat{G} \hat{c}_{k j}^{\dag}|\Psi_0\rangle
\end{align}
where $\hat{G}=\frac{1}{\omega-\hat{H}}$
and
\begin{align*}
\hat{c}_{k i}^{\dag}=\frac{1}{\sqrt{L}}\sum_{l}^{L} e^{-i k\cdot(R_l+r_i)}\hat{c}_{l i }^{\dag} \, ; \ \ \hat{c}_{k i}=\frac{1}{\sqrt{L}}\sum_{l}^{L} e^{ik\cdot(R_l+r_i)}\hat{c}_{l i }
\end{align*}
with $R_l$ the  lattice vectors (L $\rightarrow \infty$) and $r_i$ the atomic positions inside the unit cell.
(These relations hold in any spin sector and we have therefore intentionally omitted the spin index).

In the following we will adopt a many body technique to calculate the one-particle dressed Green's function based on the CPT~\cite{Senechal,Senechal3}. This method shares with  other Quantum Cluster formalisms  the basic idea of
approximating the effects of correlations in the infinite lattice  with those on a finite-size cluster. Different Quantum Cluster approaches differ for the strategy adopted to embed the cluster in the continuum and to express  the lattice Green's function --or the corresponding self-energy-- in terms of the cluster one. The common starting point is  the choice of the $M$-site cluster used to \emph{tile} the extended lattice.

In CPT  the Green's function (\ref{gij}) for the extended lattice is calculated by solving the equation
\begin{equation}\label{QC}
G_{i j}(k,   \omega)=G^c_{ i j}(  \omega)+ \sum_{i'}^M B_{i i'}(k,  \omega) G_{i' j}(k,   \omega).
\end{equation}
Here $G^c_{ i j}$ is the cluster Green's function in the local basis obtained by exact diagonalization of the interacting hamiltonian for the finite cluster; we separately solve the   problem for  N, N-1 and N+1 electrons and
express the cluster Green's function in the Lehmann representation at real frequencies.  The matrix $B_{i i'}(k,\omega)$ is given by
\begin{align*}
B_{i i'}(k,\omega)= \sum_l^L e^{i k\cdot R_l} \sum_{i''}^M G^c_{i i''}(\omega) t_{i'' 0, i' l}(s)
\end{align*}
where $t_{i''0 ,i' l}$ is the hopping term between site $i'$ and $i''$ belonging to different clusters.

Eq.~\eqref{QC} is  solved  by a $M\times M$  matrix inversion at each $k$ and $\omega$.
A second $M\times M$ matrix inversion is needed to obtain the topological hamiltonian according to eq.~\eqref{htopo}. The diagonalization of the  topological hamiltonian is then required to obtain the  eigenvectors to be used for the calculation of    $\mathbb{Z}_2$ according to~\eqref{z2}.
It is worth recalling that  the
eigenvalues of $h_{topo}$ in principle have nothing to do with the quasi-particle excitation energies: only the topological information is  encoded in $G_{i j}(k, 0)$,  but the full Green's function is needed to calculate quasi-particle spectral functions
\begin{align}
\label{akn}
A(k,\omega) = \frac{1}{\pi}\sum_n {\rm Im}\,  G(k,n,\omega)
\end{align}
where
 \begin{align*}
G(k,n,\omega) =  \frac{1}{M}\sum_{i i'} e^{-i k \cdot(r_i-r_{i'})}\alpha^{n*}_i(k) \alpha^n_{i'}(k)  G_{i i'}(k,\omega)
 \end{align*}
with $n$ the band index and $\alpha^n_i(k)$ the eigenstate coefficients obtained by the single-particle band calculation.\cite{ManghiMnO}
In the next section, analyzing in the detail all the information that can be deduced from the explicit calculation of the interacting Green's function,  we will also be able to investigate more closely the relations between the eigenstates of the topological hamiltonian and the quasi-particle energies.

\section{Results}
\label{sec3}

\begin{figure}[htbp]
\includegraphics[width=8.5cm]{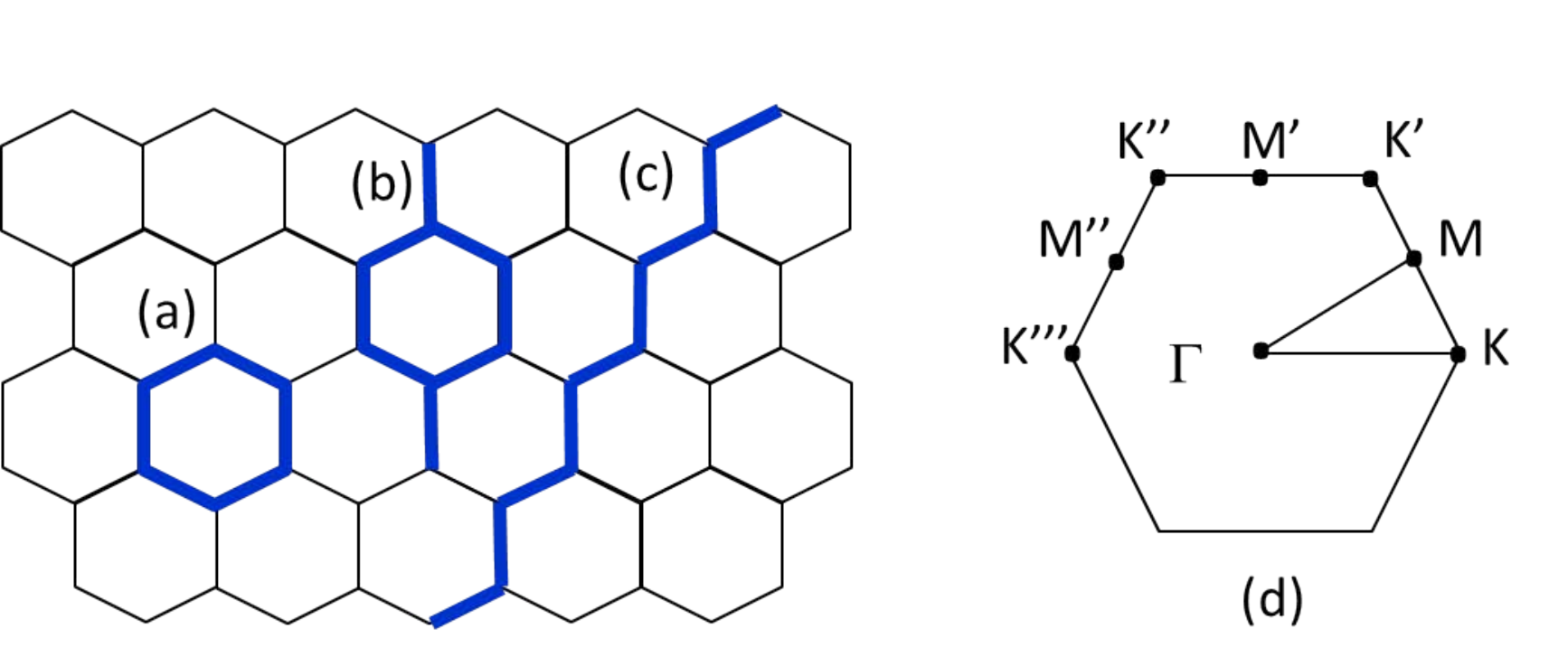}
  \caption{\label{geometry} (Color online) 6-site tiling (a) and 8-site tiling (b) of the 2D honeycomb lattice.  (c) 10 site chain cluster used to tile the 1D zigzag ribbon. This cluster is vertically repeated to describe ribbons of increasing width. (d) 2D Brillouine zone.}
\end{figure}

We have used the CPT formalism to calculate the dressed Green's function of the Kane-Mele-Hubbard model  spanning a whole set of spin-orbit couplings $t_{KM}$ and $U$ parameters. For the 2D   honeycomb lattice the 6-site cluster (Fig.~\ref{geometry} (a)) commonly used in Quantum Cluster calculations~\cite{Yu_ribbon_TI,PhysRevB.85.205102,PhysRevB.86.045105,Liebsch} has been adopted. In order to check the role of cluster size and geometry we have also considered the 8-site  cluster of Fig. \ref{geometry} (b). Both clusters represent a \emph{tiling} for the honeycomb lattice but with very different  cluster symmetries. Obviously this has no influence in the non-interacting case where either  the ``natural" 2-site unit cell  or any larger unit cell (4, 6, 8  sites etc.) produce the same band structure. This is no more so  if the e-e interaction is switched on:  in any Quantum Cluster approach  where the extended system is described as a periodic repetition  of correlated units, the translation periodicity  is only partially restored (it is preserved only  at the superlattice level). This inevitably affects the quasi-particle band structure and for the 8-site tiling gives rise to a wrong $k$-dispersion. This appears quite clearly by comparing  spectral functions (cfr. eq.~\eqref{akn})  obtained with 6-site and 8-site tilings at $k$-points along the border of the 2D Brillouine zone. In particular, for  the 6-site tiling  the quasi-particle energies display the correct symmetry, and  energies at any $k$-point and its rotated counterpart---$\overrightarrow{k}$ and $R \overrightarrow{k}$, with $R$ a point group rotation---coincide (see fig. 2 (a), (c) ). This well-known basic rule is violated  for the 8-site tiling   and the quasi-particle energies   at $K$, $K'''$ do not coincide  with the values at $K'$ and $K''$  (see fig. 2 (b), (d) ).  Indeed the gap closes down around $K'$ and $K''$ but not at $K$ and $K'''$.  This is due to the fact that the 8-site tiling has a preferred direction so that the dispersions along $K-K'$  and $K'-K''$ are different. 

 The dependence  on the cluster's size and symmetry of the quasiparticle band dispersion in the Kane-Mele-Hubbard model is shown here for the first time  and appears to be  crucial in order to identify the accuracy and appropriateness of the results: any band structure of a non-interacting system  violating the point group symmetries should be disregarded as wrong and unphysical;   the same should be done for  interacting systems since e-e repulsion does not  affect the lattice point group symmetry.  For this reason the semimetal   behaviour for $t_{KM}=0$  and  $U/t \leq 3.5$   that we find for the 8-site tiling  in agreement with Refs. \onlinecite{PhysRevB.85.205102,PhysRevB.90.165136}   
should  be considered an artifact  due to the wrong cluster symmetry and  not to be used to infer any real behaviour of the model system.    Only clusters that preserve the point group symmetries of the lattice should  be used~\cite{RevModPhysQC}  and this criterion restricts the choice for the 2D honeycomb lattice to the 6-site cluster.~\footnote{A larger cluster with the correct point group symmetry would contain  24 sites and would be too large for exact diagonalization}. These considerations are  quite general and  quasi-particle states, topological invariants and  phase diagrams  obtained by Quantum Cluster approaches using   tilings with the wrong symmetry (2-, 4-, 8-site clusters) \cite{PhysRevB.85.205102,PhysRevB.86.201407,PhysRevB.90.165136}  are for this reason not reliable.

\begin{figure}[htbp]
\includegraphics[width=8.75cm]{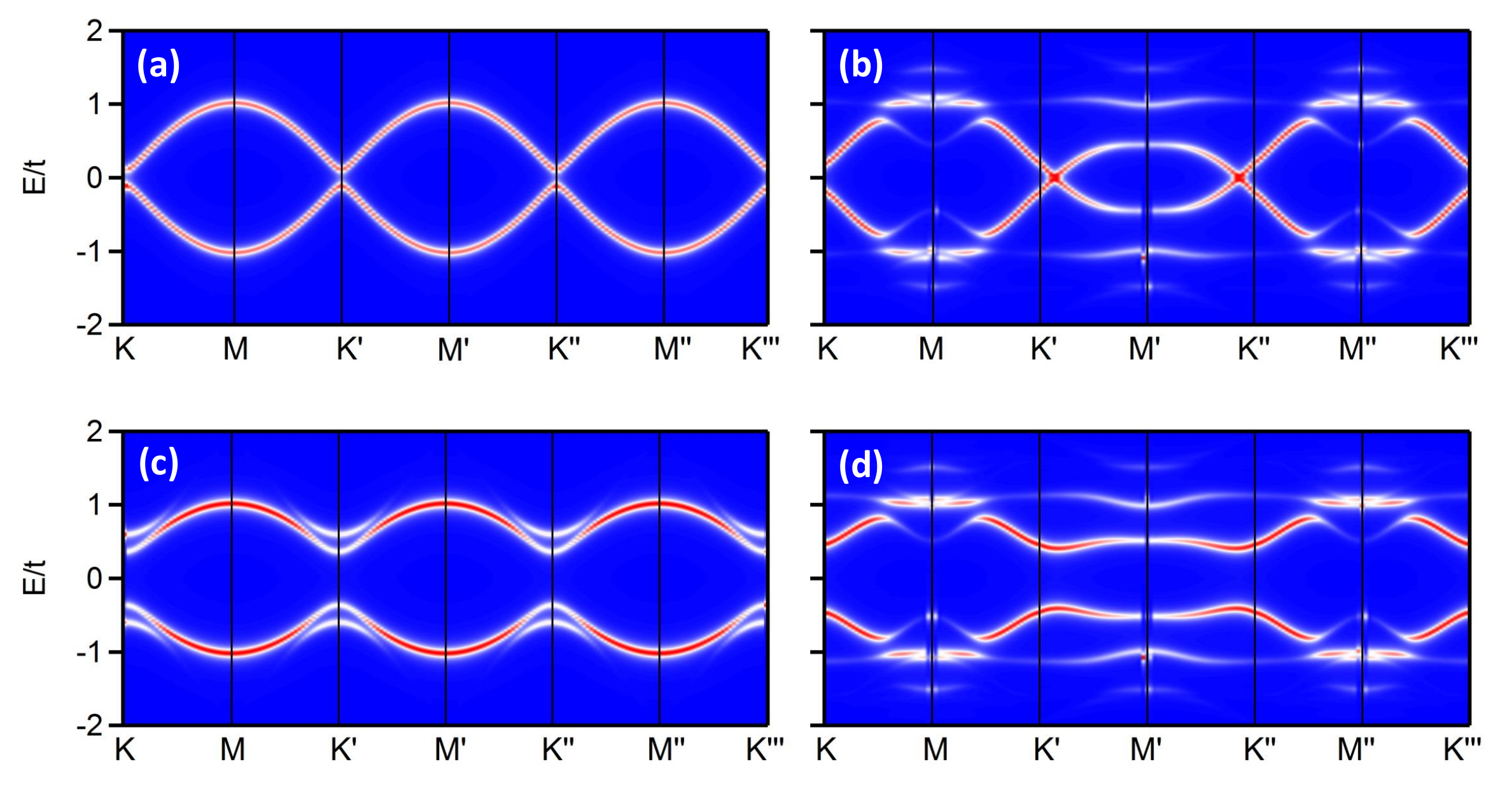}
  \caption{\label{geometry2} (Color online) Comparison between the quasi-particle band structures obtained for  6-site and 8-site tiling assuming $U/t=2$. Panel (a) and (b)  ( (c) and (d) )  show the results obtained with $t_{KM}/t=0$ ( $t_{KM}/t=0.1$) for 6-site and 8-site tiling respectively. Notice that the gapless band structure obtained for $t_{KM}/t=0$ with the 8-site tiling is a consequence of a band dispersion that  violates the rotational symmetry of the lattice.}
\end{figure}

\begin{figure}
\includegraphics[width=8.75cm]{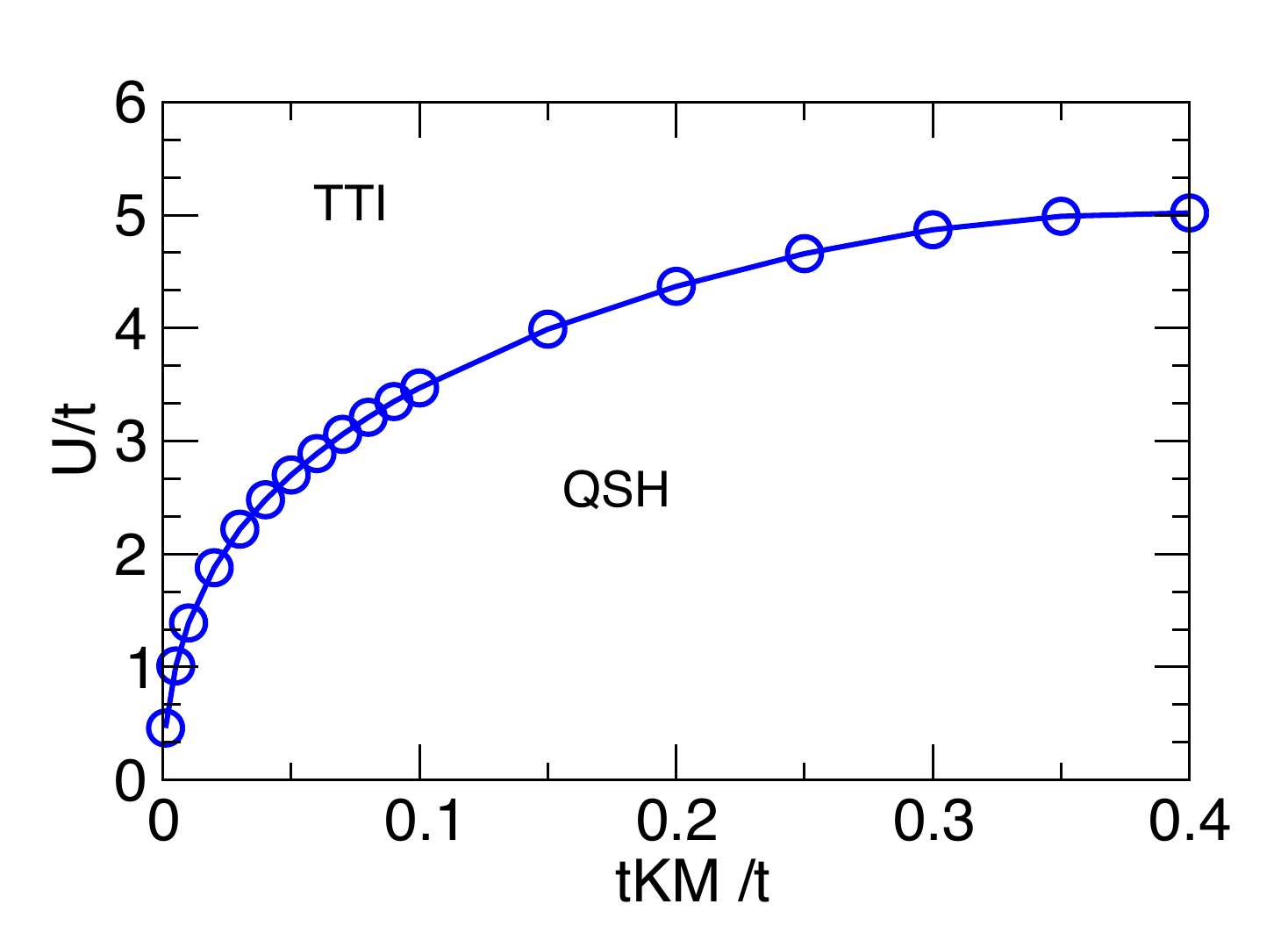}
  \caption{\label{phaseD} (Color online) $U-t_{KM}$ phase diagram of the half filled Kane-Mele-Hubbard model. The two regimes, QSH and TTI correspond  to different values of the $\mathbb{Z}_2$ invariant ($\Delta=1$ and $\Delta=0$ respectively). Open dots correspond to the parameter values where the calculation has been done, the continuous line is a guide for the eye.}
\end{figure}

We focus now on the  Green's function at $\omega=0$ and on the topological properties that can be deduced from it. As discussed in the previous section, the key quantity is the  dressed Green's function expressed in a Bloch basis (eqs.~\eqref{gij} and~\eqref{QC}) and the corresponding topological hamiltonian $(h_{topo})_{ij}=-G^{-1}_{i j}(k,\omega=0)$.
The eigenvalue problem associated to $h_{topo}$  --a   $6 \times 6$ matrix diagonalization at each $k$-point-- is equivalent to a standard single particle Tight-Binding  calculation for a unit cell containing 6 atomic sites, giving rise to 6 topological bands. The product over the first 3 occupied bands at the TRIM points corresponding to the 6-site cluster provides, according to eq.~\eqref{z2},   the  $\mathbb{Z}_2$  invariant. Fig.~\ref{phaseD} reports the  resulting $U-t_{KM}$ phase diagram showing the parameter range where the system behaves as either a topologically trivial insulator (TTI, $\Delta=0$) or  a Quantum Spin Hall  insulator (QSH, $\Delta=1$) \footnote{The transport properties of the present model are determined by the value of the topological invariant $\Delta$. The invariant is a discrete function, and thus the transition between TTI and QSH can be identified as a first order phase transition. This is confirmed by earlier studies of quantum phase transitions in interacting topological insulators, e.g.~\onlinecite{PhysRevB.82.115125}.}.

Few comments are in order: since we are monitoring the topological phase transition by  the $\mathbb{Z}_2$ invariant, we are implicitly assuming both time reversal and parity invariance and an adiabatic connection between QSH and TTI phase to persist in all regimes. Anti-ferromagnetism that breaks both  time-reversal and sublattice inversion symmetry is therefore excluded and we are assuming  the system to remain non-magnetic  at any $U$. In this sense the value of $\mathbb{Z}_2$ can be considered an indicator of the topological properties of the system of interest only for a parameter range that excludes antiferromagnetism.

\begin{figure}
\includegraphics[width=8.75cm]{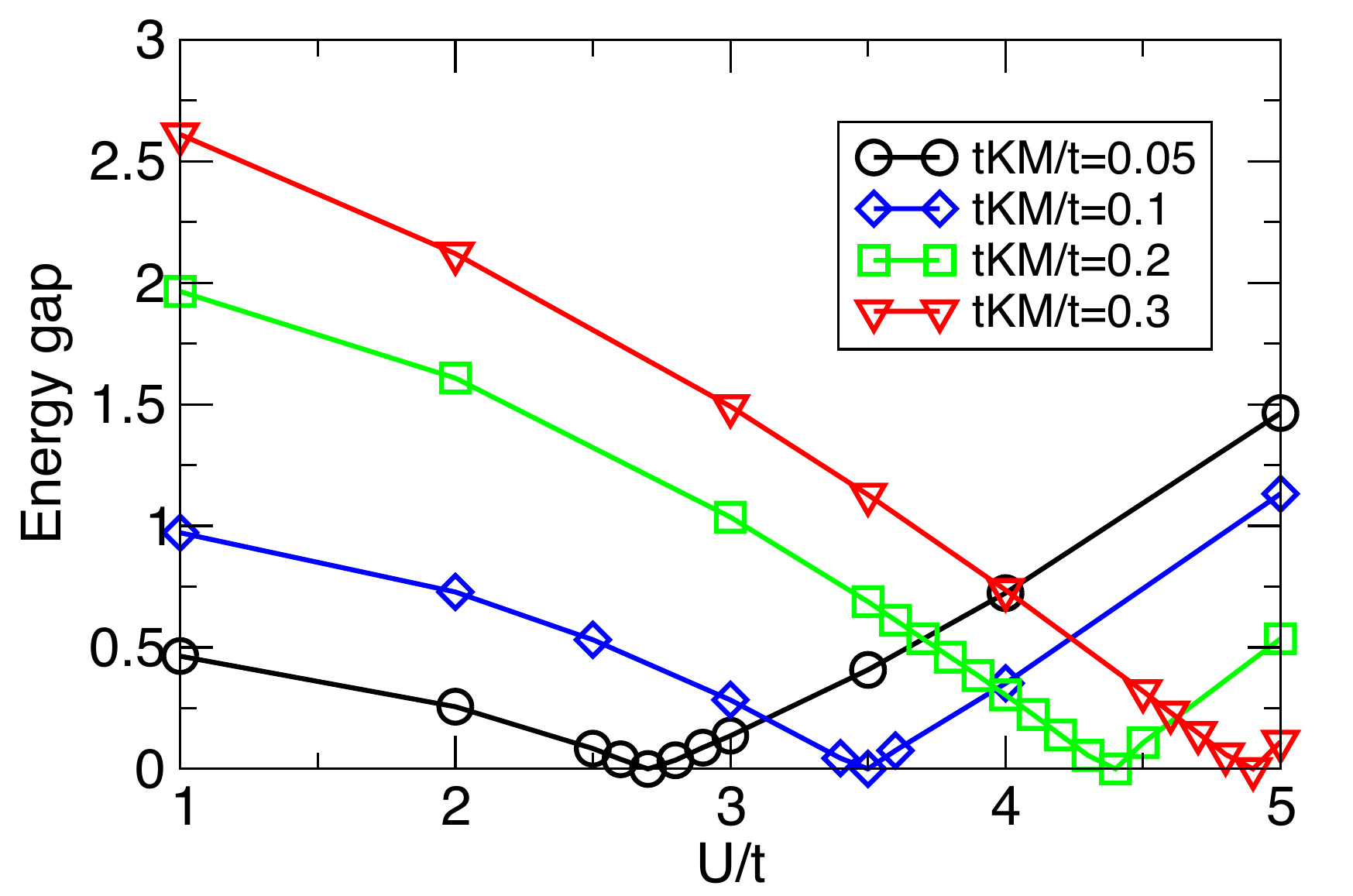}
  \caption{\label{gap} (Color online) Value of the energy gap $\Delta_{sp}$ as a function of $U/t$ for different values of the intrinsic spin-orbit  parameter $t_{KM}/t$. }
\end{figure}

The  behavior for $t_{KM}$ close to zero is worth noticing. According to our  CPT calculation, at low  $t_{KM}$  the QSH regime  does not survive the switching on of e-e interaction: a value $U \to 0$ is enough to destroy the  semi-metallic behavior at $t_{KM}=0$; see Fig.~\ref{gap}.  This is at variance  with Quantum Monte Carlo (QMC) results~\cite{Meng,sorella} that at $t_{KM}=0$  identify a semimetallic behavior up to $U/t\sim 3.5$ \footnote{The existence of a spin liquid phase between the semi-metallic and anti-ferromagnetic ones predicted  in ref.~\onlinecite{Meng} has been successively ruled out by more refined QMC calculations in ref.~\onlinecite{sorella}. }.
It has been recently shown~\cite{PhysRevB.87.205127} that the existence at $t_{KM}=0$ of an excitation gap down  to $U \to 0$ is characteristic of all Quantum Cluster schemes with the only exception of DCA. This is due to  the aforementioned violation of translational symmetry in Quantum Cluster methods such as CPT, VCA and CDMFT,  regardless of the scheme being variational or not, and independent on the details of the specific implementations (different impurity solvers, different temperatures). We stress here  again that the semimetal behaviour that is found for the Kane-Mele-Hubbard model by Quantum Cluster approaches such as  CDMFT ~\cite{PhysRevB.85.205102} and VCA ~\cite{PhysRevB.90.165136}  is actually an artifact due to the choice of clusters with wrong symmetry. The only Quantum Cluster approach that is able to reproduce a semimetal behaviour at finite $U$   is DCA.   
DCA preserves  translation symmetry and has been shown to describe better the small $U$ regime; it becomes however less accurate at large $U$ values where it overemphasizes the semimetallic  behavior of the honeycomb lattice.~\cite{PhysRevB.87.205127} In this sense DCA and the other Quantum Cluster approaches can be considered as complementary and it would be interesting to compare their results also in terms of parity invariants.

\begin{figure}[htbp]
\includegraphics[width=8.75cm]{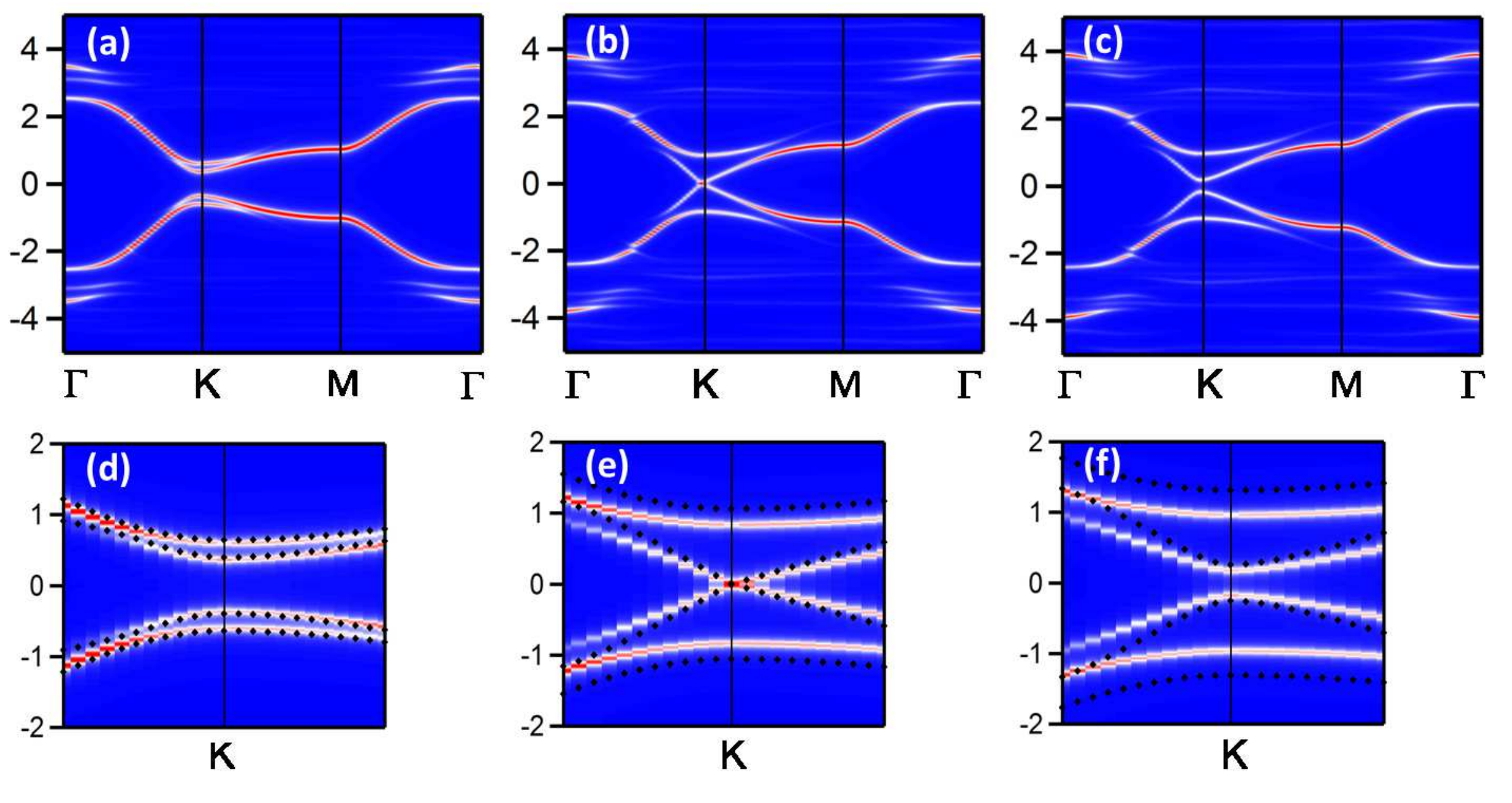}
  \caption{\label{spectral} (Color online) Spectral functions of the 2D honeycomb lattice for $t_{KM}/t=0.1$ and $U/t=2$ (a), $U/t=3.5$ (b), $U/t=4$ (c). A zoom of the energy region around the Fermi energy is shown for the three cases in panel (d)-(f) respectively. The corresponding eigenvalues of $h_{topo}$ are superimposed as black dots.}
\end{figure}

By calculating spectral functions in the same parameter range we observe that at the transition points the single particle excitation gap $\Delta _{sp}$, namely the minimum energy separation between hole and particle excitations, closes down. The possibility for  e-e interaction to induce a metallic behavior in a band insulator has been recently analysed by DMFT \cite{PhysRevLett.97.046403,PhysRevB.80.155116} and QMC calculations \cite{PhysRevLett.98.046403}; here we observe that, in agreement with previous results \cite{Yu_ribbon_TI,PhysRevB.85.205102}, the same effect occurs in the  honeycomb lattice made semiconducting by intrinsic spin-orbit interaction.  Fig. \ref{gap} shows  the behavior of  $\Delta _{sp}$ at different values of $t_{KM}/t$  as a function of $U/t$.   Fig. \ref{spectral} (a-c) shows as an example the quasi-particle band structure obtained for  $t_{KM}/t=0.1$ and for $U/t =2$  (QSH regime), $U/t =3.5 $ (transition  point from QSH to TTI, the gap closes down)  and $U/t =4 $ (TTI regime).

\begin{figure}
\includegraphics[width=8.75cm]{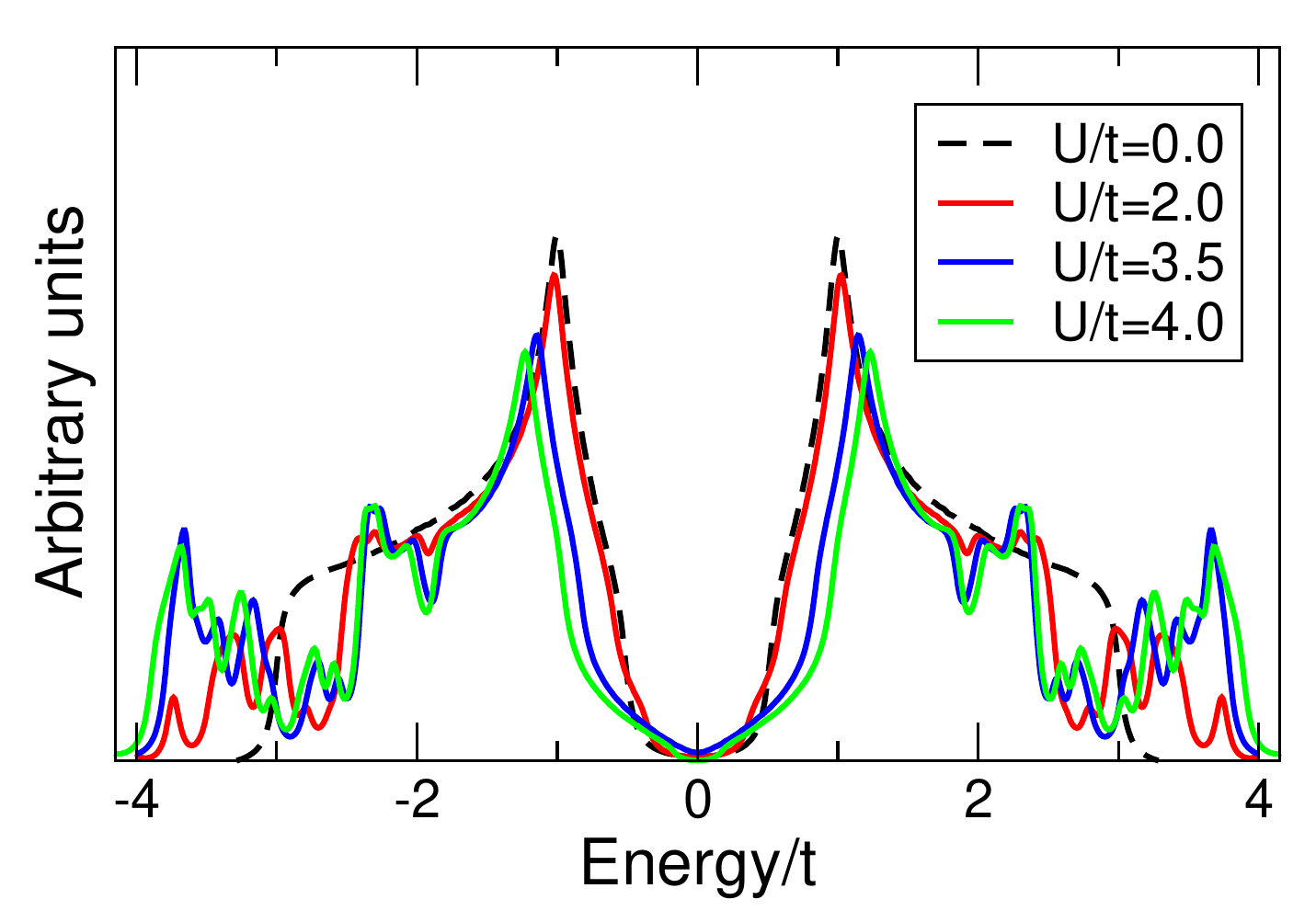}
  \caption{\label{DOSC} (Color online) Density of states  of the half filled Kane-Mele-Hubbard model for $t_{KM}/t=0.1$ at $U/t=2$ (red), $3.5 $ (blue), $4$ (green) compared with the non interacting results (dashed line). Satellite structures appear below and above the valence and conduction band respectively and the band width is reduced, an effect that is more evident for larger $U$.}
\end{figure}

Other effects are  due to the e-e correlation, namely a band width reduction and the appearance of satellite structures below (above)  valence (conduction) band.   These effects are more clearly seen by looking at the density of  states (DOS) obtained as the sum of the spectral functions over a large sample of $k$-points (Fig.~\ref{DOSC}).

Even if the eigenvalues  of $h_{topo}$  cannot be  identified with excitation energies,
they  exhibit a  behavior similar to  the quasi-particle energies. In particular the same gap closure appears in $h_{topo}$ eigenvalues at the transition points. This is shown in Fig.~\ref{spectral} (d-f) where a zoom of the quasi-particle band structure around the point $K$ is compared with the eigenvalues $\epsilon_{n k \uparrow}$ of $h_{topo}$.

We may then conclude, in agreement with QMC calculations \cite{PhysRevB.87.205101,PhysRevB.87.121113}, that a change in the $\mathbb{Z}_2$  invariant is  associated to the closure of both the single-particle excitation gap and of the energy separation between filled and empty states of $h_{topo}$: in strict analogy with  the non-interacting case, a change of topological regime of  the interacting systems is  associated to a gap closure followed by a gap inversion in the fictitious band structure associated to $h_{topo}$.

\begin{figure}[htbp]
\includegraphics[width=8.75cm]{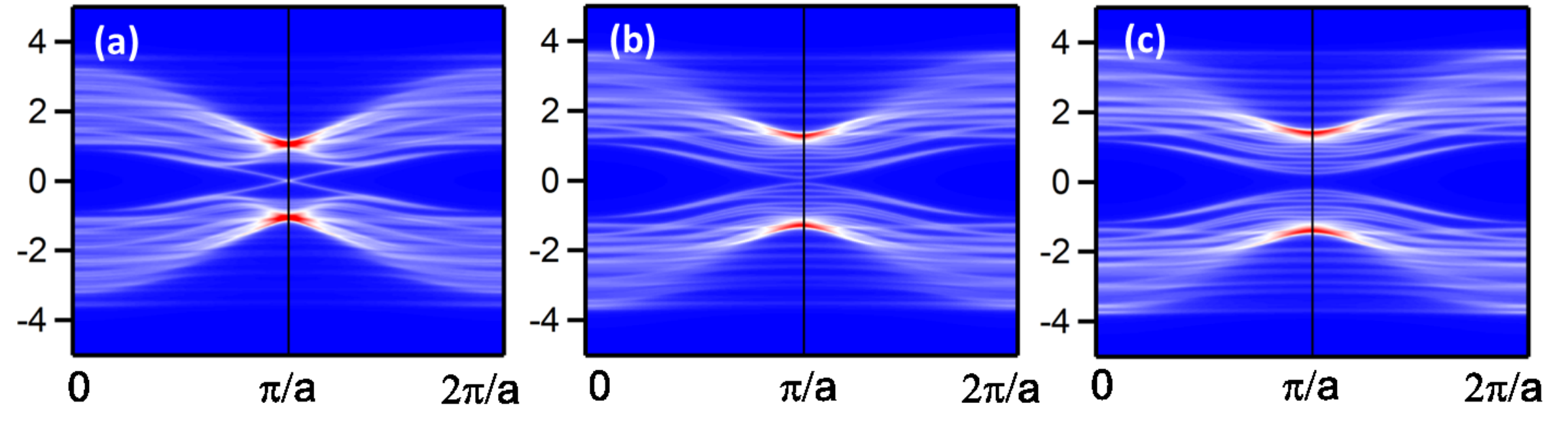}
  \caption{\label{ribbon} (Color online) Spectral functions of a zigzag  honeycomb ribbon for $t_{KM}/t=0.1$ and $U/t=2$ (a), $U/t=3.5$ (b), $U/t=4$ (c).  The ribbon width corresponds to 30 sites per cell.}
\end{figure}

According to the bulk-boundary correspondence, 1D non-interacting systems should exhibit gapless edge states once the 2D system enters the QSH regime. In the presence of e-e interaction this may not be true and  a gap may open in edge states  before the time-reversal Z2 invariant switches off~\cite{PhysRevB.84.205121}. We have calculated within CPT the spectral functions  for a  honeycomb ribbon with zigzag termination using the tiling shown in Fig. \ref{geometry} (c). For any given value of $t_{KM}$ we have systematically found that gapless edge states persist up to a critical value of $U$ that coincides with the one previously identified as the transition point from QSH to TTI  regime in the 2D system. This is shown in Fig. \ref{ribbon} for $t_{KM}/t= 0.1$. Here we notice that  at critical value $U/t=3.5$ a tiny gap  exists between filled and empty states. We have checked that, increasing the ribbon width, this gap becomes smaller and smaller,  and we may then attribute it to  the finite width of the ribbon.

In conclusion we have studied the  topological properties of the Kane-Mele-Hubbard model by explicitly calculating the  Green's function- based topological invariant. The  CPT scheme,  using  Bloch sums as a basis set,  naturally leads to the topological hamiltonian matrix  that enters in the computation of the $\mathbb{Z}_2$ topological invariant. The approach gives direct access to the dressed Green's function at real frequencies, avoiding the problem of analytic continuation, and does not require the extraction of a self-energy.
We have shown that the interplay between the Hubbard interaction and the intrinsic spin-orbit coupling is coherently described by  the values of $\mathbb{Z}_2$  invariant and by  2D and 1D quasi-particle  energies (gap closing at the transitions, edge states in the QSH phase). We have discussed the importance of the cluster symmetry and the effects of   the lack of full translation symmetry typical of CPT and of most  Quantum Cluster approaches. Comments on the limits of applicability of the method are also provided.


\begin{thebibliography}{53}%
\makeatletter
\providecommand \@ifxundefined [1]{%
 \@ifx{#1\undefined}
}%
\providecommand \@ifnum [1]{%
 \ifnum #1\expandafter \@firstoftwo
 \else \expandafter \@secondoftwo
 \fi
}%
\providecommand \@ifx [1]{%
 \ifx #1\expandafter \@firstoftwo
 \else \expandafter \@secondoftwo
 \fi
}%
\providecommand \natexlab [1]{#1}%
\providecommand \enquote  [1]{``#1''}%
\providecommand \bibnamefont  [1]{#1}%
\providecommand \bibfnamefont [1]{#1}%
\providecommand \citenamefont [1]{#1}%
\providecommand \href@noop [0]{\@secondoftwo}%
\providecommand \href [0]{\begingroup \@sanitize@url \@href}%
\providecommand \@href[1]{\@@startlink{#1}\@@href}%
\providecommand \@@href[1]{\endgroup#1\@@endlink}%
\providecommand \@sanitize@url [0]{\catcode `\\12\catcode `\$12\catcode
  `\&12\catcode `\#12\catcode `\^12\catcode `\_12\catcode `\%12\relax}%
\providecommand \@@startlink[1]{}%
\providecommand \@@endlink[0]{}%
\providecommand \url  [0]{\begingroup\@sanitize@url \@url }%
\providecommand \@url [1]{\endgroup\@href {#1}{\urlprefix }}%
\providecommand \urlprefix  [0]{URL }%
\providecommand \Eprint [0]{\href }%
\providecommand \doibase [0]{http://dx.doi.org/}%
\providecommand \selectlanguage [0]{\@gobble}%
\providecommand \bibinfo  [0]{\@secondoftwo}%
\providecommand \bibfield  [0]{\@secondoftwo}%
\providecommand \translation [1]{[#1]}%
\providecommand \BibitemOpen [0]{}%
\providecommand \bibitemStop [0]{}%
\providecommand \bibitemNoStop [0]{.\EOS\space}%
\providecommand \EOS [0]{\spacefactor3000\relax}%
\providecommand \BibitemShut  [1]{\csname bibitem#1\endcsname}%
\let\auto@bib@innerbib\@empty
\bibitem [{\citenamefont {Hasan}\ and\ \citenamefont
  {Kane}(2010)}]{RevModPhys.82.3045}%
  \BibitemOpen
  \bibfield  {author} {\bibinfo {author} {\bibfnamefont {M.~Z.}\ \bibnamefont
  {Hasan}}\ and\ \bibinfo {author} {\bibfnamefont {C.~L.}\ \bibnamefont
  {Kane}},\ }\href {\doibase 10.1103/RevModPhys.82.3045} {\bibfield  {journal}
  {\bibinfo  {journal} {Rev. Mod. Phys.}\ }\textbf {\bibinfo {volume} {82}},\
  \bibinfo {pages} {3045} (\bibinfo {year} {2010})}\BibitemShut {NoStop}%
\bibitem [{\citenamefont {Ando}(2013)}]{Ando}%
  \BibitemOpen
  \bibfield  {author} {\bibinfo {author} {\bibfnamefont {Y.}~\bibnamefont
  {Ando}},\ }\href {http://dx.doi.org/10.7566/JPSJ.82.102001} {\bibfield
  {journal} {\bibinfo  {journal} {J. Phys. Soc. Japan}\ }\textbf {\bibinfo
  {volume} {82}},\ \bibinfo {pages} {102001} (\bibinfo {year}
  {2013})}\BibitemShut {NoStop}%
\bibitem [{\citenamefont {Qi}\ \emph {et~al.}(2006)\citenamefont {Qi},
  \citenamefont {Wu},\ and\ \citenamefont {Zhang}}]{PhysRevB.74.045125}%
  \BibitemOpen
  \bibfield  {author} {\bibinfo {author} {\bibfnamefont {X.-L.}\ \bibnamefont
  {Qi}}, \bibinfo {author} {\bibfnamefont {Y.-S.}\ \bibnamefont {Wu}}, \ and\
  \bibinfo {author} {\bibfnamefont {S.-C.}\ \bibnamefont {Zhang}},\ }\href
  {\doibase 10.1103/PhysRevB.74.045125} {\bibfield  {journal} {\bibinfo
  {journal} {Phys. Rev. B}\ }\textbf {\bibinfo {volume} {74}},\ \bibinfo
  {pages} {045125} (\bibinfo {year} {2006})}\BibitemShut {NoStop}%
\bibitem [{\citenamefont {Xu}\ and\ \citenamefont
  {Moore}(2006)}]{PhysRevB.73.045322}%
  \BibitemOpen
  \bibfield  {author} {\bibinfo {author} {\bibfnamefont {C.}~\bibnamefont
  {Xu}}\ and\ \bibinfo {author} {\bibfnamefont {J.~E.}\ \bibnamefont {Moore}},\
  }\href {\doibase 10.1103/PhysRevB.73.045322} {\bibfield  {journal} {\bibinfo
  {journal} {Phys. Rev. B}\ }\textbf {\bibinfo {volume} {73}},\ \bibinfo
  {pages} {045322} (\bibinfo {year} {2006})}\BibitemShut {NoStop}%
\bibitem [{\citenamefont {Wu}\ \emph {et~al.}(2006)\citenamefont {Wu},
  \citenamefont {Bernevig},\ and\ \citenamefont
  {Zhang}}]{PhysRevLett.96.106401}%
  \BibitemOpen
  \bibfield  {author} {\bibinfo {author} {\bibfnamefont {C.}~\bibnamefont
  {Wu}}, \bibinfo {author} {\bibfnamefont {B.~A.}\ \bibnamefont {Bernevig}}, \
  and\ \bibinfo {author} {\bibfnamefont {S.-C.}\ \bibnamefont {Zhang}},\ }\href
  {\doibase 10.1103/PhysRevLett.96.106401} {\bibfield  {journal} {\bibinfo
  {journal} {Phys. Rev. Lett.}\ }\textbf {\bibinfo {volume} {96}},\ \bibinfo
  {pages} {106401} (\bibinfo {year} {2006})}\BibitemShut {NoStop}%
\bibitem [{\citenamefont {Kane}\ and\ \citenamefont
  {Mele}(2005)}]{PhysRevLett.95.146802}%
  \BibitemOpen
  \bibfield  {author} {\bibinfo {author} {\bibfnamefont {C.~L.}\ \bibnamefont
  {Kane}}\ and\ \bibinfo {author} {\bibfnamefont {E.~J.}\ \bibnamefont
  {Mele}},\ }\href {\doibase 10.1103/PhysRevLett.95.146802} {\bibfield
  {journal} {\bibinfo  {journal} {Phys. Rev. Lett.}\ }\textbf {\bibinfo
  {volume} {95}},\ \bibinfo {pages} {146802} (\bibinfo {year}
  {2005})}\BibitemShut {NoStop}%
\bibitem [{\citenamefont {Fu}\ and\ \citenamefont
  {Kane}(2006)}]{PhysRevB.74.195312}%
  \BibitemOpen
  \bibfield  {author} {\bibinfo {author} {\bibfnamefont {L.}~\bibnamefont
  {Fu}}\ and\ \bibinfo {author} {\bibfnamefont {C.~L.}\ \bibnamefont {Kane}},\
  }\href {\doibase 10.1103/PhysRevB.74.195312} {\bibfield  {journal} {\bibinfo
  {journal} {Phys. Rev. B}\ }\textbf {\bibinfo {volume} {74}},\ \bibinfo
  {pages} {195312} (\bibinfo {year} {2006})}\BibitemShut {NoStop}%
\bibitem [{\citenamefont {Fu}\ and\ \citenamefont
  {Kane}(2007)}]{PhysRevB.76.045302}%
  \BibitemOpen
  \bibfield  {author} {\bibinfo {author} {\bibfnamefont {L.}~\bibnamefont
  {Fu}}\ and\ \bibinfo {author} {\bibfnamefont {C.~L.}\ \bibnamefont {Kane}},\
  }\href {\doibase 10.1103/PhysRevB.76.045302} {\bibfield  {journal} {\bibinfo
  {journal} {Phys. Rev. B}\ }\textbf {\bibinfo {volume} {76}},\ \bibinfo
  {pages} {045302} (\bibinfo {year} {2007})}\BibitemShut {NoStop}%
\bibitem [{\citenamefont {Thouless}\ \emph {et~al.}(1982)\citenamefont
  {Thouless}, \citenamefont {Kohmoto}, \citenamefont {Nightingale},\ and\
  \citenamefont {den Nijs}}]{Thouless:1982zz}%
  \BibitemOpen
  \bibfield  {author} {\bibinfo {author} {\bibfnamefont {D.}~\bibnamefont
  {Thouless}}, \bibinfo {author} {\bibfnamefont {M.}~\bibnamefont {Kohmoto}},
  \bibinfo {author} {\bibfnamefont {M.}~\bibnamefont {Nightingale}}, \ and\
  \bibinfo {author} {\bibfnamefont {M.}~\bibnamefont {den Nijs}},\ }\href
  {\doibase 10.1103/PhysRevLett.49.405} {\bibfield  {journal} {\bibinfo
  {journal} {Phys.Rev.Lett.}\ }\textbf {\bibinfo {volume} {49}},\ \bibinfo
  {pages} {405} (\bibinfo {year} {1982})}\BibitemShut {NoStop}%
\bibitem [{\citenamefont {Niu}\ \emph {et~al.}(1985)\citenamefont {Niu},
  \citenamefont {Thouless},\ and\ \citenamefont {Wu}}]{PhysRevB.31.3372}%
  \BibitemOpen
  \bibfield  {author} {\bibinfo {author} {\bibfnamefont {Q.}~\bibnamefont
  {Niu}}, \bibinfo {author} {\bibfnamefont {D.~J.}\ \bibnamefont {Thouless}}, \
  and\ \bibinfo {author} {\bibfnamefont {Y.-S.}\ \bibnamefont {Wu}},\ }\href
  {\doibase 10.1103/PhysRevB.31.3372} {\bibfield  {journal} {\bibinfo
  {journal} {Phys. Rev. B}\ }\textbf {\bibinfo {volume} {31}},\ \bibinfo
  {pages} {3372} (\bibinfo {year} {1985})}\BibitemShut {NoStop}%
\bibitem [{\citenamefont {Wang}\ \emph {et~al.}(2012)\citenamefont {Wang},
  \citenamefont {Qi},\ and\ \citenamefont {Zhang}}]{PhysRevB.85.165126}%
  \BibitemOpen
  \bibfield  {author} {\bibinfo {author} {\bibfnamefont {Z.}~\bibnamefont
  {Wang}}, \bibinfo {author} {\bibfnamefont {X.-L.}\ \bibnamefont {Qi}}, \ and\
  \bibinfo {author} {\bibfnamefont {S.-C.}\ \bibnamefont {Zhang}},\ }\href
  {\doibase 10.1103/PhysRevB.85.165126} {\bibfield  {journal} {\bibinfo
  {journal} {Phys. Rev. B}\ }\textbf {\bibinfo {volume} {85}},\ \bibinfo
  {pages} {165126} (\bibinfo {year} {2012})}\BibitemShut {NoStop}%
\bibitem [{\citenamefont {Wang}\ and\ \citenamefont
  {Zhang}(2012)}]{PhysRevX.2.031008}%
  \BibitemOpen
  \bibfield  {author} {\bibinfo {author} {\bibfnamefont {Z.}~\bibnamefont
  {Wang}}\ and\ \bibinfo {author} {\bibfnamefont {S.-C.}\ \bibnamefont
  {Zhang}},\ }\href {\doibase 10.1103/PhysRevX.2.031008} {\bibfield  {journal}
  {\bibinfo  {journal} {Phys. Rev. X}\ }\textbf {\bibinfo {volume} {2}},\
  \bibinfo {pages} {031008} (\bibinfo {year} {2012})}\BibitemShut {NoStop}%
\bibitem [{\citenamefont {Gurarie}(2011)}]{PhysRevB.83.085426}%
  \BibitemOpen
  \bibfield  {author} {\bibinfo {author} {\bibfnamefont {V.}~\bibnamefont
  {Gurarie}},\ }\href {\doibase 10.1103/PhysRevB.83.085426} {\bibfield
  {journal} {\bibinfo  {journal} {Phys. Rev. B}\ }\textbf {\bibinfo {volume}
  {83}},\ \bibinfo {pages} {085426} (\bibinfo {year} {2011})}\BibitemShut
  {NoStop}%
\bibitem [{\citenamefont {Hohenadler}\ and\ \citenamefont
  {Assaad}(2013)}]{Assaadrev}%
  \BibitemOpen
  \bibfield  {author} {\bibinfo {author} {\bibfnamefont {M.}~\bibnamefont
  {Hohenadler}}\ and\ \bibinfo {author} {\bibfnamefont {F.~F.}\ \bibnamefont
  {Assaad}},\ }\href@noop {} {\bibfield  {journal} {\bibinfo  {journal}
  {Journal of Physics: Condensed Matter}\ }\textbf {\bibinfo {volume} {25}},\
  \bibinfo {pages} {143201} (\bibinfo {year} {2013})}\BibitemShut {NoStop}%
\bibitem [{\citenamefont {Meng}\ \emph {et~al.}(2014)\citenamefont {Meng},
  \citenamefont {Hung},\ and\ \citenamefont
  {Lang}}]{doi:10.1142/S0217984914300014}%
  \BibitemOpen
  \bibfield  {author} {\bibinfo {author} {\bibfnamefont {Z.~Y.}\ \bibnamefont
  {Meng}}, \bibinfo {author} {\bibfnamefont {H.-H.}\ \bibnamefont {Hung}}, \
  and\ \bibinfo {author} {\bibfnamefont {T.~C.}\ \bibnamefont {Lang}},\ }\href
  {\doibase 10.1142/S0217984914300014} {\bibfield  {journal} {\bibinfo
  {journal} {Modern Physics Letters B}\ }\textbf {\bibinfo {volume} {28}},\
  \bibinfo {pages} {1430001} (\bibinfo {year} {2014})}\BibitemShut {NoStop}%
\bibitem [{\citenamefont {Werner}\ and\ \citenamefont
  {Assaad}(2013)}]{PhysRevB.88.035113}%
  \BibitemOpen
  \bibfield  {author} {\bibinfo {author} {\bibfnamefont {J.}~\bibnamefont
  {Werner}}\ and\ \bibinfo {author} {\bibfnamefont {F.~F.}\ \bibnamefont
  {Assaad}},\ }\href {\doibase 10.1103/PhysRevB.88.035113} {\bibfield
  {journal} {\bibinfo  {journal} {Phys. Rev. B}\ }\textbf {\bibinfo {volume}
  {88}},\ \bibinfo {pages} {035113} (\bibinfo {year} {2013})}\BibitemShut
  {NoStop}%
\bibitem [{\citenamefont {Deng}\ \emph {et~al.}(2013)\citenamefont {Deng},
  \citenamefont {Haule},\ and\ \citenamefont
  {Kotliar}}]{PhysRevLett.111.176404}%
  \BibitemOpen
  \bibfield  {author} {\bibinfo {author} {\bibfnamefont {X.}~\bibnamefont
  {Deng}}, \bibinfo {author} {\bibfnamefont {K.}~\bibnamefont {Haule}}, \ and\
  \bibinfo {author} {\bibfnamefont {G.}~\bibnamefont {Kotliar}},\ }\href
  {\doibase 10.1103/PhysRevLett.111.176404} {\bibfield  {journal} {\bibinfo
  {journal} {Phys. Rev. Lett.}\ }\textbf {\bibinfo {volume} {111}},\ \bibinfo
  {pages} {176404} (\bibinfo {year} {2013})}\BibitemShut {NoStop}%
\bibitem [{\citenamefont {Lu}\ \emph {et~al.}(2013)\citenamefont {Lu},
  \citenamefont {Zhao}, \citenamefont {Weng}, \citenamefont {Fang},\ and\
  \citenamefont {Dai}}]{PhysRevLett.110.096401}%
  \BibitemOpen
  \bibfield  {author} {\bibinfo {author} {\bibfnamefont {F.}~\bibnamefont
  {Lu}}, \bibinfo {author} {\bibfnamefont {J.}~\bibnamefont {Zhao}}, \bibinfo
  {author} {\bibfnamefont {H.}~\bibnamefont {Weng}}, \bibinfo {author}
  {\bibfnamefont {Z.}~\bibnamefont {Fang}}, \ and\ \bibinfo {author}
  {\bibfnamefont {X.}~\bibnamefont {Dai}},\ }\href {\doibase
  10.1103/PhysRevLett.110.096401} {\bibfield  {journal} {\bibinfo  {journal}
  {Phys. Rev. Lett.}\ }\textbf {\bibinfo {volume} {110}},\ \bibinfo {pages}
  {096401} (\bibinfo {year} {2013})}\BibitemShut {NoStop}%
\bibitem [{\citenamefont {Yoshida}\ \emph {et~al.}(2013)\citenamefont
  {Yoshida}, \citenamefont {Peters}, \citenamefont {Fujimoto},\ and\
  \citenamefont {Kawakami}}]{PhysRevB.87.085134}%
  \BibitemOpen
  \bibfield  {author} {\bibinfo {author} {\bibfnamefont {T.}~\bibnamefont
  {Yoshida}}, \bibinfo {author} {\bibfnamefont {R.}~\bibnamefont {Peters}},
  \bibinfo {author} {\bibfnamefont {S.}~\bibnamefont {Fujimoto}}, \ and\
  \bibinfo {author} {\bibfnamefont {N.}~\bibnamefont {Kawakami}},\ }\href
  {\doibase 10.1103/PhysRevB.87.085134} {\bibfield  {journal} {\bibinfo
  {journal} {Phys. Rev. B}\ }\textbf {\bibinfo {volume} {87}},\ \bibinfo
  {pages} {085134} (\bibinfo {year} {2013})}\BibitemShut {NoStop}%
\bibitem [{\citenamefont {Lang}\ \emph {et~al.}(2013)\citenamefont {Lang},
  \citenamefont {Essin}, \citenamefont {Gurarie},\ and\ \citenamefont
  {Wessel}}]{PhysRevB.87.205101}%
  \BibitemOpen
  \bibfield  {author} {\bibinfo {author} {\bibfnamefont {T.~C.}\ \bibnamefont
  {Lang}}, \bibinfo {author} {\bibfnamefont {A.~M.}\ \bibnamefont {Essin}},
  \bibinfo {author} {\bibfnamefont {V.}~\bibnamefont {Gurarie}}, \ and\
  \bibinfo {author} {\bibfnamefont {S.}~\bibnamefont {Wessel}},\ }\href
  {\doibase 10.1103/PhysRevB.87.205101} {\bibfield  {journal} {\bibinfo
  {journal} {Phys. Rev. B}\ }\textbf {\bibinfo {volume} {87}},\ \bibinfo
  {pages} {205101} (\bibinfo {year} {2013})}\BibitemShut {NoStop}%
\bibitem [{\citenamefont {Maier}\ \emph {et~al.}(2005)\citenamefont {Maier},
  \citenamefont {Jarrell}, \citenamefont {Pruschke},\ and\ \citenamefont
  {Hettler}}]{RevModPhysQC}%
  \BibitemOpen
  \bibfield  {author} {\bibinfo {author} {\bibfnamefont {T.}~\bibnamefont
  {Maier}}, \bibinfo {author} {\bibfnamefont {M.}~\bibnamefont {Jarrell}},
  \bibinfo {author} {\bibfnamefont {T.}~\bibnamefont {Pruschke}}, \ and\
  \bibinfo {author} {\bibfnamefont {M.~H.}\ \bibnamefont {Hettler}},\ }\href
  {\doibase 10.1103/RevModPhys.77.1027} {\bibfield  {journal} {\bibinfo
  {journal} {Rev. Mod. Phys.}\ }\textbf {\bibinfo {volume} {77}},\ \bibinfo
  {pages} {1027} (\bibinfo {year} {2005})}\BibitemShut {NoStop}%
\bibitem [{\citenamefont {S\'en\'echal}\ \emph {et~al.}(2000)\citenamefont
  {S\'en\'echal}, \citenamefont {Perez},\ and\ \citenamefont
  {Pioro-Ladri\`ere}}]{Senechal}%
  \BibitemOpen
  \bibfield  {author} {\bibinfo {author} {\bibfnamefont {D.}~\bibnamefont
  {S\'en\'echal}}, \bibinfo {author} {\bibfnamefont {D.}~\bibnamefont {Perez}},
  \ and\ \bibinfo {author} {\bibfnamefont {M.}~\bibnamefont
  {Pioro-Ladri\`ere}},\ }\href {\doibase 10.1103/PhysRevLett.84.522} {\bibfield
   {journal} {\bibinfo  {journal} {Phys. Rev. Lett.}\ }\textbf {\bibinfo
  {volume} {84}},\ \bibinfo {pages} {522} (\bibinfo {year} {2000})}\BibitemShut
  {NoStop}%
\bibitem [{\citenamefont {S{\'e}n{\'e}chal}(2012)}]{Senechal3}%
  \BibitemOpen
  \bibfield  {author} {\bibinfo {author} {\bibfnamefont {D.}~\bibnamefont
  {S{\'e}n{\'e}chal}},\ }\enquote {\bibinfo {title} {Cluster perturbation
  theory},}\ in\ \href@noop {} {\emph {\bibinfo {booktitle} {Theoretical
  methods for Strongly Correlated Systems}}},\ \bibinfo {series} {Springer
  Series in Solid-State Sciences}, Vol.\ \bibinfo {volume} {171}\ (\bibinfo
  {publisher} {Springer},\ \bibinfo {year} {2012})\ Chap.~\bibinfo {chapter}
  {8}, pp.\ \bibinfo {pages} {237--269}\BibitemShut {NoStop}%
\bibitem [{\citenamefont {Hettler}\ \emph {et~al.}(1998)\citenamefont
  {Hettler}, \citenamefont {Tahvildar-Zadeh}, \citenamefont {Jarrell},
  \citenamefont {Pruschke},\ and\ \citenamefont {Krishnamurthy}}]{DCA}%
  \BibitemOpen
  \bibfield  {author} {\bibinfo {author} {\bibfnamefont {M.~H.}\ \bibnamefont
  {Hettler}}, \bibinfo {author} {\bibfnamefont {A.~N.}\ \bibnamefont
  {Tahvildar-Zadeh}}, \bibinfo {author} {\bibfnamefont {M.}~\bibnamefont
  {Jarrell}}, \bibinfo {author} {\bibfnamefont {T.}~\bibnamefont {Pruschke}}, \
  and\ \bibinfo {author} {\bibfnamefont {H.~R.}\ \bibnamefont
  {Krishnamurthy}},\ }\href {\doibase 10.1103/PhysRevB.58.R7475} {\bibfield
  {journal} {\bibinfo  {journal} {Phys. Rev. B}\ }\textbf {\bibinfo {volume}
  {58}},\ \bibinfo {pages} {R7475} (\bibinfo {year} {1998})}\BibitemShut
  {NoStop}%
\bibitem [{\citenamefont {Potthoff}\ \emph {et~al.}(2003)\citenamefont
  {Potthoff}, \citenamefont {Aichhorn},\ and\ \citenamefont {Dahnken}}]{VCA}%
  \BibitemOpen
  \bibfield  {author} {\bibinfo {author} {\bibfnamefont {M.}~\bibnamefont
  {Potthoff}}, \bibinfo {author} {\bibfnamefont {M.}~\bibnamefont {Aichhorn}},
  \ and\ \bibinfo {author} {\bibfnamefont {C.}~\bibnamefont {Dahnken}},\ }\href
  {\doibase 10.1103/PhysRevLett.91.206402} {\bibfield  {journal} {\bibinfo
  {journal} {Phys. Rev. Lett.}\ }\textbf {\bibinfo {volume} {91}},\ \bibinfo
  {pages} {206402} (\bibinfo {year} {2003})}\BibitemShut {NoStop}%
\bibitem [{\citenamefont {Kancharla}\ \emph {et~al.}(2008)\citenamefont
  {Kancharla}, \citenamefont {Kyung}, \citenamefont {S\'en\'echal},
  \citenamefont {Civelli}, \citenamefont {Capone}, \citenamefont {Kotliar},\
  and\ \citenamefont {Tremblay}}]{Senechal_CDMFT}%
  \BibitemOpen
  \bibfield  {author} {\bibinfo {author} {\bibfnamefont {S.~S.}\ \bibnamefont
  {Kancharla}}, \bibinfo {author} {\bibfnamefont {B.}~\bibnamefont {Kyung}},
  \bibinfo {author} {\bibfnamefont {D.}~\bibnamefont {S\'en\'echal}}, \bibinfo
  {author} {\bibfnamefont {M.}~\bibnamefont {Civelli}}, \bibinfo {author}
  {\bibfnamefont {M.}~\bibnamefont {Capone}}, \bibinfo {author} {\bibfnamefont
  {G.}~\bibnamefont {Kotliar}}, \ and\ \bibinfo {author} {\bibfnamefont
  {A.-M.~S.}\ \bibnamefont {Tremblay}},\ }\href {\doibase
  10.1103/PhysRevB.77.184516} {\bibfield  {journal} {\bibinfo  {journal} {Phys.
  Rev. B}\ }\textbf {\bibinfo {volume} {77}},\ \bibinfo {pages} {184516}
  (\bibinfo {year} {2008})}\BibitemShut {NoStop}%
\bibitem [{\citenamefont {Potthoff}(2003)}]{SFA}%
  \BibitemOpen
  \bibfield  {author} {\bibinfo {author} {\bibfnamefont {M.}~\bibnamefont
  {Potthoff}},\ }\href@noop {} {\bibfield  {journal} {\bibinfo  {journal} {Eur.
  Phys. J. B}\ }\textbf {\bibinfo {volume} {32}},\ \bibinfo {pages} {245110}
  (\bibinfo {year} {2003})}\BibitemShut {NoStop}%
\bibitem [{\citenamefont {Eder}(2008)}]{Eder}%
  \BibitemOpen
  \bibfield  {author} {\bibinfo {author} {\bibfnamefont {R.}~\bibnamefont
  {Eder}},\ }\href {\doibase 10.1103/PhysRevB.78.115111} {\bibfield  {journal}
  {\bibinfo  {journal} {Phys. Rev. B}\ }\textbf {\bibinfo {volume} {78}},\
  \bibinfo {pages} {115111} (\bibinfo {year} {2008})}\BibitemShut {NoStop}%
\bibitem [{\citenamefont {Manghi}(2014)}]{ManghiMnO}%
  \BibitemOpen
  \bibfield  {author} {\bibinfo {author} {\bibfnamefont {F.}~\bibnamefont
  {Manghi}},\ }\href {http://stacks.iop.org/0953-8984/26/i=1/a=015602}
  {\bibfield  {journal} {\bibinfo  {journal} {Journal of Physics: Condensed
  Matter}\ }\textbf {\bibinfo {volume} {26}},\ \bibinfo {pages} {015602}
  (\bibinfo {year} {2014})}\BibitemShut {NoStop}%
\bibitem [{\citenamefont {Rachel}\ and\ \citenamefont
  {Le~Hur}(2010)}]{PhysRevB.82.075106}%
  \BibitemOpen
  \bibfield  {author} {\bibinfo {author} {\bibfnamefont {S.}~\bibnamefont
  {Rachel}}\ and\ \bibinfo {author} {\bibfnamefont {K.}~\bibnamefont
  {Le~Hur}},\ }\href {\doibase 10.1103/PhysRevB.82.075106} {\bibfield
  {journal} {\bibinfo  {journal} {Phys. Rev. B}\ }\textbf {\bibinfo {volume}
  {82}},\ \bibinfo {pages} {075106} (\bibinfo {year} {2010})}\BibitemShut
  {NoStop}%
\bibitem [{\citenamefont {Zheng}\ \emph {et~al.}(2011)\citenamefont {Zheng},
  \citenamefont {Zhang},\ and\ \citenamefont {Wu}}]{PhysRevB.84.205121}%
  \BibitemOpen
  \bibfield  {author} {\bibinfo {author} {\bibfnamefont {D.}~\bibnamefont
  {Zheng}}, \bibinfo {author} {\bibfnamefont {G.-M.}\ \bibnamefont {Zhang}}, \
  and\ \bibinfo {author} {\bibfnamefont {C.}~\bibnamefont {Wu}},\ }\href
  {\doibase 10.1103/PhysRevB.84.205121} {\bibfield  {journal} {\bibinfo
  {journal} {Phys. Rev. B}\ }\textbf {\bibinfo {volume} {84}},\ \bibinfo
  {pages} {205121} (\bibinfo {year} {2011})}\BibitemShut {NoStop}%
\bibitem [{\citenamefont {Hohenadler}\ \emph {et~al.}(2011)\citenamefont
  {Hohenadler}, \citenamefont {Lang},\ and\ \citenamefont
  {Assaad}}]{PhysRevLett.106.100403}%
  \BibitemOpen
  \bibfield  {author} {\bibinfo {author} {\bibfnamefont {M.}~\bibnamefont
  {Hohenadler}}, \bibinfo {author} {\bibfnamefont {T.~C.}\ \bibnamefont
  {Lang}}, \ and\ \bibinfo {author} {\bibfnamefont {F.~F.}\ \bibnamefont
  {Assaad}},\ }\href {\doibase 10.1103/PhysRevLett.106.100403} {\bibfield
  {journal} {\bibinfo  {journal} {Phys. Rev. Lett.}\ }\textbf {\bibinfo
  {volume} {106}},\ \bibinfo {pages} {100403} (\bibinfo {year}
  {2011})}\BibitemShut {NoStop}%
\bibitem [{\citenamefont {Hohenadler}\ \emph {et~al.}(2012)\citenamefont
  {Hohenadler}, \citenamefont {Meng}, \citenamefont {Lang}, \citenamefont
  {Wessel}, \citenamefont {Muramatsu},\ and\ \citenamefont
  {Assaad}}]{PhysRevB.85.115132}%
  \BibitemOpen
  \bibfield  {author} {\bibinfo {author} {\bibfnamefont {M.}~\bibnamefont
  {Hohenadler}}, \bibinfo {author} {\bibfnamefont {Z.~Y.}\ \bibnamefont
  {Meng}}, \bibinfo {author} {\bibfnamefont {T.~C.}\ \bibnamefont {Lang}},
  \bibinfo {author} {\bibfnamefont {S.}~\bibnamefont {Wessel}}, \bibinfo
  {author} {\bibfnamefont {A.}~\bibnamefont {Muramatsu}}, \ and\ \bibinfo
  {author} {\bibfnamefont {F.~F.}\ \bibnamefont {Assaad}},\ }\href {\doibase
  10.1103/PhysRevB.85.115132} {\bibfield  {journal} {\bibinfo  {journal} {Phys.
  Rev. B}\ }\textbf {\bibinfo {volume} {85}},\ \bibinfo {pages} {115132}
  (\bibinfo {year} {2012})}\BibitemShut {NoStop}%
\bibitem [{\citenamefont {Wang}\ \emph {et~al.}(2010)\citenamefont {Wang},
  \citenamefont {Qi},\ and\ \citenamefont {Zhang}}]{PhysRevLett.105.256803}%
  \BibitemOpen
  \bibfield  {author} {\bibinfo {author} {\bibfnamefont {Z.}~\bibnamefont
  {Wang}}, \bibinfo {author} {\bibfnamefont {X.-L.}\ \bibnamefont {Qi}}, \ and\
  \bibinfo {author} {\bibfnamefont {S.-C.}\ \bibnamefont {Zhang}},\ }\href
  {\doibase 10.1103/PhysRevLett.105.256803} {\bibfield  {journal} {\bibinfo
  {journal} {Phys. Rev. Lett.}\ }\textbf {\bibinfo {volume} {105}},\ \bibinfo
  {pages} {256803} (\bibinfo {year} {2010})}\BibitemShut {NoStop}%
\bibitem [{\citenamefont {Wang}\ and\ \citenamefont
  {Yan}(2013)}]{0953-8984-25-15-155601}%
  \BibitemOpen
  \bibfield  {author} {\bibinfo {author} {\bibfnamefont {Z.}~\bibnamefont
  {Wang}}\ and\ \bibinfo {author} {\bibfnamefont {B.}~\bibnamefont {Yan}},\
  }\href {http://stacks.iop.org/0953-8984/25/i=15/a=155601} {\bibfield
  {journal} {\bibinfo  {journal} {Journal of Physics: Condensed Matter}\
  }\textbf {\bibinfo {volume} {25}},\ \bibinfo {pages} {155601} (\bibinfo
  {year} {2013})}\BibitemShut {NoStop}%
\bibitem [{\citenamefont {Haldane}(1988)}]{PhysRevLett.61.2015}%
  \BibitemOpen
  \bibfield  {author} {\bibinfo {author} {\bibfnamefont {F.~D.~M.}\
  \bibnamefont {Haldane}},\ }\href {\doibase 10.1103/PhysRevLett.61.2015}
  {\bibfield  {journal} {\bibinfo  {journal} {Phys. Rev. Lett.}\ }\textbf
  {\bibinfo {volume} {61}},\ \bibinfo {pages} {2015} (\bibinfo {year}
  {1988})}\BibitemShut {NoStop}%
\bibitem [{\citenamefont {Yu}\ \emph {et~al.}(2011)\citenamefont {Yu},
  \citenamefont {Xie},\ and\ \citenamefont {Li}}]{Yu_ribbon_TI}%
  \BibitemOpen
  \bibfield  {author} {\bibinfo {author} {\bibfnamefont {S.-L.}\ \bibnamefont
  {Yu}}, \bibinfo {author} {\bibfnamefont {X.~C.}\ \bibnamefont {Xie}}, \ and\
  \bibinfo {author} {\bibfnamefont {J.-X.}\ \bibnamefont {Li}},\ }\href
  {\doibase 10.1103/PhysRevLett.107.010401} {\bibfield  {journal} {\bibinfo
  {journal} {Phys. Rev. Lett.}\ }\textbf {\bibinfo {volume} {107}},\ \bibinfo
  {pages} {010401} (\bibinfo {year} {2011})}\BibitemShut {NoStop}%
\bibitem [{\citenamefont {Wu}\ \emph {et~al.}(2012)\citenamefont {Wu},
  \citenamefont {Rachel}, \citenamefont {Liu},\ and\ \citenamefont
  {Le~Hur}}]{PhysRevB.85.205102}%
  \BibitemOpen
  \bibfield  {author} {\bibinfo {author} {\bibfnamefont {W.}~\bibnamefont
  {Wu}}, \bibinfo {author} {\bibfnamefont {S.}~\bibnamefont {Rachel}}, \bibinfo
  {author} {\bibfnamefont {W.-M.}\ \bibnamefont {Liu}}, \ and\ \bibinfo
  {author} {\bibfnamefont {K.}~\bibnamefont {Le~Hur}},\ }\href {\doibase
  10.1103/PhysRevB.85.205102} {\bibfield  {journal} {\bibinfo  {journal} {Phys.
  Rev. B}\ }\textbf {\bibinfo {volume} {85}},\ \bibinfo {pages} {205102}
  (\bibinfo {year} {2012})}\BibitemShut {NoStop}%
\bibitem [{\citenamefont {He}\ and\ \citenamefont
  {Lu}(2012)}]{PhysRevB.86.045105}%
  \BibitemOpen
  \bibfield  {author} {\bibinfo {author} {\bibfnamefont {R.-Q.}\ \bibnamefont
  {He}}\ and\ \bibinfo {author} {\bibfnamefont {Z.-Y.}\ \bibnamefont {Lu}},\
  }\href {\doibase 10.1103/PhysRevB.86.045105} {\bibfield  {journal} {\bibinfo
  {journal} {Phys. Rev. B}\ }\textbf {\bibinfo {volume} {86}},\ \bibinfo
  {pages} {045105} (\bibinfo {year} {2012})}\BibitemShut {NoStop}%
\bibitem [{\citenamefont {Liebsch}(2011)}]{Liebsch}%
  \BibitemOpen
  \bibfield  {author} {\bibinfo {author} {\bibfnamefont {A.}~\bibnamefont
  {Liebsch}},\ }\href {\doibase 10.1103/PhysRevB.83.035113} {\bibfield
  {journal} {\bibinfo  {journal} {Phys. Rev. B}\ }\textbf {\bibinfo {volume}
  {83}},\ \bibinfo {pages} {035113} (\bibinfo {year} {2011})}\BibitemShut
  {NoStop}%
  \bibitem [{Note1()}]{Note1}%
  \BibitemOpen
  \bibinfo {note} {A larger cluster with the correct point group symmetry would
  contain 24 sites and would be too large for exact
  diagonalization}\BibitemShut {NoStop}%
\bibitem [{\citenamefont {Budich}\ \emph {et~al.}(2012)\citenamefont {Budich},
  \citenamefont {Thomale}, \citenamefont {Li}, \citenamefont {Laubach},\ and\
  \citenamefont {Zhang}}]{PhysRevB.86.201407}%
  \BibitemOpen
  \bibfield  {author} {\bibinfo {author} {\bibfnamefont {J.~C.}\ \bibnamefont
  {Budich}}, \bibinfo {author} {\bibfnamefont {R.}~\bibnamefont {Thomale}},
  \bibinfo {author} {\bibfnamefont {G.}~\bibnamefont {Li}}, \bibinfo {author}
  {\bibfnamefont {M.}~\bibnamefont {Laubach}}, \ and\ \bibinfo {author}
  {\bibfnamefont {S.-C.}\ \bibnamefont {Zhang}},\ }\href {\doibase
  10.1103/PhysRevB.86.201407} {\bibfield  {journal} {\bibinfo  {journal} {Phys.
  Rev. B}\ }\textbf {\bibinfo {volume} {86}},\ \bibinfo {pages} {201407}
  (\bibinfo {year} {2012})}\BibitemShut {NoStop}%
\bibitem [{\citenamefont {Laubach}\ \emph {et~al.}(2014)\citenamefont
  {Laubach}, \citenamefont {Reuther}, \citenamefont {Thomale},\ and\
  \citenamefont {Rachel}}]{PhysRevB.90.165136}%
  \BibitemOpen
  \bibfield  {author} {\bibinfo {author} {\bibfnamefont {M.}~\bibnamefont
  {Laubach}}, \bibinfo {author} {\bibfnamefont {J.}~\bibnamefont {Reuther}},
  \bibinfo {author} {\bibfnamefont {R.}~\bibnamefont {Thomale}}, \ and\
  \bibinfo {author} {\bibfnamefont {S.}~\bibnamefont {Rachel}},\ }\href
  {\doibase 10.1103/PhysRevB.90.165136} {\bibfield  {journal} {\bibinfo
  {journal} {Phys. Rev. B}\ }\textbf {\bibinfo {volume} {90}},\ \bibinfo
  {pages} {165136} (\bibinfo {year} {2014})}\BibitemShut {NoStop}%
\bibitem [{Note2()}]{Note2}%
  \BibitemOpen
  \bibinfo {note} {The transport properties of the present model are determined
  by the value of the topological invariant $\Delta $. The invariant is a
  discrete function, and thus the transition between TTI and QSH can be
  identified as a first order phase transition. This is confirmed by earlier
  studies of quantum phase transitions in interacting topological insulators,
  e.g.~\protect \rev@citealpnum {PhysRevB.82.115125}.}\BibitemShut {Stop}%
\bibitem [{\citenamefont {{Meng}}\ \emph {et~al.}(2010)\citenamefont {{Meng}},
  \citenamefont {{Lang}}, \citenamefont {{Wessel}}, \citenamefont {{Assaad}},\
  and\ \citenamefont {{Muramatsu}}}]{Meng}%
  \BibitemOpen
  \bibfield  {author} {\bibinfo {author} {\bibfnamefont {Z.~Y.}\ \bibnamefont
  {{Meng}}}, \bibinfo {author} {\bibfnamefont {T.~C.}\ \bibnamefont {{Lang}}},
  \bibinfo {author} {\bibfnamefont {S.}~\bibnamefont {{Wessel}}}, \bibinfo
  {author} {\bibfnamefont {F.~F.}\ \bibnamefont {{Assaad}}}, \ and\ \bibinfo
  {author} {\bibfnamefont {A.}~\bibnamefont {{Muramatsu}}},\ }\href {\doibase
  10.1038/nature08942} {\bibfield  {journal} {\bibinfo  {journal} {Nature}\
  }\textbf {\bibinfo {volume} {464}},\ \bibinfo {pages} {847} (\bibinfo {year}
  {2010})} \BibitemShut {NoStop}%
\bibitem [{\citenamefont {Sorella}\ \emph {et~al.}(2012)\citenamefont
  {Sorella}, \citenamefont {Otsuka},\ and\ \citenamefont {Yunoki}}]{sorella}%
  \BibitemOpen
  \bibfield  {author} {\bibinfo {author} {\bibfnamefont {S.}~\bibnamefont
  {Sorella}}, \bibinfo {author} {\bibfnamefont {Y.}~\bibnamefont {Otsuka}}, \
  and\ \bibinfo {author} {\bibfnamefont {S.}~\bibnamefont {Yunoki}},\ }\href
  {http://www.nature.com/srep/2012/121218/srep00992/full/srep00992.html}
  {\bibfield  {journal} {\bibinfo  {journal} {Sci. Rep.}\ }\textbf {\bibinfo
  {volume} {2}},\ \bibinfo {pages} {992} (\bibinfo {year} {2012})}\BibitemShut
  {NoStop}%
\bibitem [{Note3()}]{Note3}%
  \BibitemOpen
  \bibinfo {note} {The existence of a spin liquid phase between the
  semi-metallic and anti-ferromagnetic ones predicted in ref.~\protect
  \rev@citealpnum {Meng} has been successively ruled out by more refined QMC
  calculations in ref.~\protect \rev@citealpnum {sorella}.}\BibitemShut {Stop}%
\bibitem [{\citenamefont {Liebsch}\ and\ \citenamefont
  {Wu}(2013)}]{PhysRevB.87.205127}%
  \BibitemOpen
  \bibfield  {author} {\bibinfo {author} {\bibfnamefont {A.}~\bibnamefont
  {Liebsch}}\ and\ \bibinfo {author} {\bibfnamefont {W.}~\bibnamefont {Wu}},\
  }\href {\doibase 10.1103/PhysRevB.87.205127} {\bibfield  {journal} {\bibinfo
  {journal} {Phys. Rev. B}\ }\textbf {\bibinfo {volume} {87}},\ \bibinfo
  {pages} {205127} (\bibinfo {year} {2013})}\BibitemShut {NoStop}%
\bibitem [{\citenamefont {Garg}\ \emph {et~al.}(2006)\citenamefont {Garg},
  \citenamefont {Krishnamurthy},\ and\ \citenamefont
  {Randeria}}]{PhysRevLett.97.046403}%
  \BibitemOpen
  \bibfield  {author} {\bibinfo {author} {\bibfnamefont {A.}~\bibnamefont
  {Garg}}, \bibinfo {author} {\bibfnamefont {H.~R.}\ \bibnamefont
  {Krishnamurthy}}, \ and\ \bibinfo {author} {\bibfnamefont {M.}~\bibnamefont
  {Randeria}},\ }\href {\doibase 10.1103/PhysRevLett.97.046403} {\bibfield
  {journal} {\bibinfo  {journal} {Phys. Rev. Lett.}\ }\textbf {\bibinfo
  {volume} {97}},\ \bibinfo {pages} {046403} (\bibinfo {year}
  {2006})}\BibitemShut {NoStop}%
\bibitem [{\citenamefont {Sentef}\ \emph {et~al.}(2009)\citenamefont {Sentef},
  \citenamefont {Kune\ifmmode~\check{s}\else \v{s}\fi{}}, \citenamefont
  {Werner},\ and\ \citenamefont {Kampf}}]{PhysRevB.80.155116}%
  \BibitemOpen
  \bibfield  {author} {\bibinfo {author} {\bibfnamefont {M.}~\bibnamefont
  {Sentef}}, \bibinfo {author} {\bibfnamefont {J.}~\bibnamefont
  {Kune\ifmmode~\check{s}\else \v{s}\fi{}}}, \bibinfo {author} {\bibfnamefont
  {P.}~\bibnamefont {Werner}}, \ and\ \bibinfo {author} {\bibfnamefont {A.~P.}\
  \bibnamefont {Kampf}},\ }\href {\doibase 10.1103/PhysRevB.80.155116}
  {\bibfield  {journal} {\bibinfo  {journal} {Phys. Rev. B}\ }\textbf {\bibinfo
  {volume} {80}},\ \bibinfo {pages} {155116} (\bibinfo {year}
  {2009})}\BibitemShut {NoStop}%
\bibitem [{\citenamefont {Paris}\ \emph {et~al.}(2007)\citenamefont {Paris},
  \citenamefont {Bouadim}, \citenamefont {Hebert}, \citenamefont {Batrouni},\
  and\ \citenamefont {Scalettar}}]{PhysRevLett.98.046403}%
  \BibitemOpen
  \bibfield  {author} {\bibinfo {author} {\bibfnamefont {N.}~\bibnamefont
  {Paris}}, \bibinfo {author} {\bibfnamefont {K.}~\bibnamefont {Bouadim}},
  \bibinfo {author} {\bibfnamefont {F.}~\bibnamefont {Hebert}}, \bibinfo
  {author} {\bibfnamefont {G.~G.}\ \bibnamefont {Batrouni}}, \ and\ \bibinfo
  {author} {\bibfnamefont {R.~T.}\ \bibnamefont {Scalettar}},\ }\href {\doibase
  10.1103/PhysRevLett.98.046403} {\bibfield  {journal} {\bibinfo  {journal}
  {Phys. Rev. Lett.}\ }\textbf {\bibinfo {volume} {98}},\ \bibinfo {pages}
  {046403} (\bibinfo {year} {2007})}\BibitemShut {NoStop}%
\bibitem [{\citenamefont {Hung}\ \emph {et~al.}(2013)\citenamefont {Hung},
  \citenamefont {Wang}, \citenamefont {Gu},\ and\ \citenamefont
  {Fiete}}]{PhysRevB.87.121113}%
  \BibitemOpen
  \bibfield  {author} {\bibinfo {author} {\bibfnamefont {H.-H.}\ \bibnamefont
  {Hung}}, \bibinfo {author} {\bibfnamefont {L.}~\bibnamefont {Wang}}, \bibinfo
  {author} {\bibfnamefont {Z.-C.}\ \bibnamefont {Gu}}, \ and\ \bibinfo {author}
  {\bibfnamefont {G.~A.}\ \bibnamefont {Fiete}},\ }\href {\doibase
  10.1103/PhysRevB.87.121113} {\bibfield  {journal} {\bibinfo  {journal} {Phys.
  Rev. B}\ }\textbf {\bibinfo {volume} {87}},\ \bibinfo {pages} {121113}
  (\bibinfo {year} {2013})}\BibitemShut {NoStop}%
\bibitem [{\citenamefont {Varney}\ \emph {et~al.}(2010)\citenamefont {Varney},
  \citenamefont {Sun}, \citenamefont {Rigol},\ and\ \citenamefont
  {Galitski}}]{PhysRevB.82.115125}%
  \BibitemOpen
  \bibfield  {author} {\bibinfo {author} {\bibfnamefont {C.~N.}\ \bibnamefont
  {Varney}}, \bibinfo {author} {\bibfnamefont {K.}~\bibnamefont {Sun}},
  \bibinfo {author} {\bibfnamefont {M.}~\bibnamefont {Rigol}}, \ and\ \bibinfo
  {author} {\bibfnamefont {V.}~\bibnamefont {Galitski}},\ }\href {\doibase
  10.1103/PhysRevB.82.115125} {\bibfield  {journal} {\bibinfo  {journal} {Phys.
  Rev. B}\ }\textbf {\bibinfo {volume} {82}},\ \bibinfo {pages} {115125}
  (\bibinfo {year} {2010})}\BibitemShut {NoStop}%
\end{thebibliography}

%

\end{document}